\documentclass[usenatbib]{mn2e}
\usepackage{graphicx}
\usepackage{natbib}
\usepackage{amssymb}
\usepackage{aas_macros}
\usepackage{rotating}

\title[The structure of Andromeda II]{Deconstructing dwarf galaxies: a
Suprime-Cam survey of Andromeda II\thanks{Based on data collected at
Subaru Telescope, which is operated by the National Astronomical
Observatory of Japan}}

\author[McConnachie, Arimoto \& Irwin] {Alan W. McConnachie$^1$, Nobuo Arimoto$^{2,3}$ \& Mike Irwin$^4$\\
$^1$Department of Physics and Astronomy, University of Victoria, Victoria, B.C., V8P 1A1, Canada\\
$^2$National Astronomical Observatory of Japan, 2-21-1 Osawa, Mitaka, Tokyo 181-8588, Japan\\
$^3$Department of Astronomy, Graduate University of Advanced Studies,
  Mitaka, Tokyo 181-8588, Japan\\
$^4$Institute of Astronomy, University of Cambridge, Madingley Road, Cambridge, CB3 0HA, U.K.}

\begin{document}

\maketitle

\begin{abstract}
We present deep, sub-horizontal branch, multi-colour photometry of the
Andromeda~II dwarf spheroidal (And~II dSph) taken with the Subaru
Suprime-Cam wide field camera. We identify a red clump population in
this galaxy, the first time this feature has been detected in a M31
dSph, which are normally characterized as having no significant
intermediate age populations.  We construct radial profiles for the
various stellar populations and show that the horizontal branch has a
nearly constant density spatial distribution out to large radius,
whereas the reddest red giant branch stars are centrally concentrated
in an exponential profile. We argue that these populations trace two
distinct structural components in And~II, and show that this
assumption provides a good match to the overall radial profile of this
galaxy. The extended component dominates the stellar populations at
large radius, whereas the exponential component dominates the inner
few arcminutes. By examining colour-magnitude diagrams in these
regions, we show that the two components have very different stellar
populations; the exponential component has an average age of $\sim 7 -
10$\,Gyrs old, is relatively metal-rich ([Fe/H] $\sim -1$) but with a
significant tail to low metallicities, and possesses a red clump. The
extended component, on the other hand, is ancient ($\sim 13$\,Gyrs),
metal-poor ([Fe/H] $\sim -1.5$) with a narrower dispersion
$\sigma_{\rm [Fe/H]} \simeq 0.28$, and has a well developed blue
horizontal branch.  The extended component contains approximately
three-quarters of the light of And~II and its unusual density profile
is unique in Local Group dwarf galaxies. This suggests that its
formation and/or evolution may have been quite different to other
dwarf galaxies. The obvious chemo-dynamical complexity of And~II lends
further support to the accumulating body of evidence which shows that
the evolutionary histories of faint dSph galaxies can be every bit as
complicated as their brighter and more massive counterparts.
\end{abstract}


\section{Introduction}

The classic view of dwarf spheroidal (dSph) galaxies consisting of a
single, old, stellar population has changed drastically in recent
years, as deeper and more detailed observations of the Milky Way (MW)
satellites have been conducted. The shift has been so dramatic that
only a minority of the MW dSphs are now suspected of being composed of
a single stellar population, with most demonstrating multiple epochs
of star formation (for recent reviews, see
\citealt{grebel1997,mateo1998a,skillman2005}, and references therein). As
emphasised in \cite{grebel1997}, no dwarf galaxies in the Local Group
appear to share the same star formation history (SFH).

Most recently, various groups are independently showing that the
evolutionary complexities revealed in the SFHs of some dwarf galaxies
also extend to their global structural properties. \cite{harbeck2001}
investigated the presence of population gradients in the stellar
populations of a selection of Local Group dwarf galaxies, by comparing
the ratio of blue-to-red horizontal branch (HB) stars, and blue-to-red
red giant branch (RGB) stars.  These colour differences should reflect
changes in the ages and/or metallicities of the stellar
populations. They found evidence for gradients in six out of the nine
systems they studied. Recent spectroscopic studies of Sculptor
(\citealt{tolstoy2004}), Fornax (\citealt{battaglia2006}) and Canes
Venatici (\citealt{ibata2006}) have shown that these galaxies possess
spatially and {\it kinematically} distinct stellar populations. The
radial gradients in these systems therefore reflect the changing
contributions of the distinct stellar components as a function of
radius. However, \cite{koch2006a,koch2006b} do not detect any radial
gradients or kinematically distinct population in their spectroscopic
studies of Leo~I and Leo~II, and \cite{koch2006c} do not
spectroscopically detect a metallicity gradient in Carina despite a
strong age gradient implied by \cite{harbeck2001}. At this time, it is
therefore unclear how common all these features are, particularly
regarding kinematically distinct populations.

Less is known about the global stellar populations of the M31 dSphs
than for the MW dwarfs. However, several of the dSph galaxies studied
by \cite{harbeck2001} belong to the M31 subgroup. Hubble Space
Telescope (HST) WFPC2 imaging by \cite{dacosta1996,dacosta2000,dacosta2002}
and related programs were used to explore the presence of gradients
in these systems. \cite{harbeck2001} found that only Andromeda
(hereafter And)~I and VI in this subgroup show clear evidence for
gradients. This agrees with the analyses by Da Costa et al. on the
same data, where no radial gradient was detected over the fields
sampling And~II and III (\citealt{dacosta2000,dacosta2002}), whereas a
gradient was observed in And~I (\citealt{dacosta1996}).

\cite{mcconnachie2006b} (hereafter MI6) have recently presented global
wide-field photometry of Andromeda~I, II, III, V, VI and VII taken
with the Isaac Newton Telescope Wide Field Camera (INT~WFC), sampling
the top few magnitudes of the RGBs of these systems. By analysing the
global stellar structure of these galaxies, they find that And~II
shows a factor of two excess of stars in the central regions, above
that obtained by a simple extrapolation of the outer surface
brightness profile to small radii. MI6 suggest that this is evidence
showing And~II consists of two structural components with distinct
spatial distributions, similar to Sculptor and Fornax. If the stellar
populations of these components are different, then we might expect
radial gradients to be detected in this galaxy as the relative
contribution of stars from each component changes. This gradient would
probably act on a scale larger than the HST field observed by
\cite{dacosta2000}, otherwise it would likely have been detected by
this study.  The existence of multiple populations in And~II would be
consistent with the fact that \cite{dacosta2000} were unable to match
the HB morphology of this galaxy using a stellar population with a
single age, and implied that populations of multiple ages must be
present.

Multi-colour photometry which reaches the HB level can be a key
discriminant of structural variations in a dwarf galaxy
(eg. \citealt{harbeck2001}), and is a particularly powerful tool when
the data covers the entire projected area of the dwarf so that a
global view is obtained (eg. \citealt{tolstoy2004,battaglia2006}). In
addition, data of this type provides much stronger constraints on the
ages, metallicities and SFHs of the stellar populations in the galaxy
than can be achieved by analysis of the RGB alone. In this paper, we
present deep, global $VI$ photometry of And~II which reaches to below
the HB level, obtained as part of a survey of Local Group dwarf
galaxies with the Subaru Suprime-Cam wide field camera. We use these
data to probe the structures, ages and metallicities of the stellar
populations of this dwarf galaxy over its entire spatial extent, to a
depth equivalent to the earlier HST-WFPC2 study by \cite{dacosta2000}.

This paper is organised as follows: in Section~2, we introduce the
survey, and discuss the targets, observations and data analysis
procedure. In Section~3, we present colour -- magnitude diagrams
(CMDs) for And~II, and examine the spatial properties, ages and
metallicities of the dominant stellar populations. In Section~4, we
discuss the structure and evolution of And~II in light of these
results. Section~5 summarises. We assume a distance modulus to And~II
of $\left(m - M\right)_o = 24.07 \pm 0.06$ ($d = 652 \pm 18$\,kpc;
\citealt{mcconnachie2004a,mcconnachie2005a}). The average extinction
in the direction of And~II is $E\left(B - V\right) = 0.063$
(\citealt{schlegel1998}).

\section{The Subaru Suprime-Cam Survey}

Inevitably, the majority of our detailed information on the stellar
content and evolution of dSphs comes from observations of the MW
subgroup, although there are several notable exceptions. Environmental
effects, such as ram pressure stripping and tidal effects, are thought
to play an important role in the evolution of dwarf galaxies
(eg. \citealt{einasto1974,mayer2001a,mayer2001b}), and so dwarfs which
have evolved in different environments could possess very different
properties. For example, MI6 have shown that the dSph satellites of
M31 are a factor of $2 - 3$ times more extended than for the MW
population, showing that, in at least one respect, results derived
from the MW population do not necessarily hold for other Local Group
dSphs. Similarly, \cite{mcconnachie2006a} show that the radial
distribution of all of M31's dwarf satellites is nearly twice as
extended as that of the MW population. It is clearly of considerable
importance to obtain a complete census of nearby dwarf galaxy
properties, but it is only relatively recently that the
instrumentation has existed to probe the global stellar content of the
more distant dwarf galaxies of the Local Group.

As a follow-up to our INT WFC survey of Local Group dwarf galaxies
(\citealt{mcconnachie2004a,mcconnachie2005a}, MI6), we have undertaken
a deeper, wide-field, multi-colour photometric survey of Local Group
dwarf galaxies using the Subaru Suprime-Cam wide-field camera, with
the aim of obtaining photometry which reaches below the HB components
in each of our targets over their entire spatial extent. Data of this
type are generally lacking for the galaxies of the Local Group which
are not satellites of the MW, but hold a large amount of information
relating to their global structural properties and SFHs.  Subaru
Suprime-Cam is ideally suited to this task; it is able to reach the HB
of Local Group dwarf galaxies located beyond 500\,kpc in a reasonable
amount of time and, importantly, the large ($34$\,arcmins $\times
27$\,arcmins) field of view is sufficient to cover the entire dwarf
galaxy at this distance.

\subsection{The survey}

\begin{figure*}
  \begin{center}
    \includegraphics[angle=0, width=12.cm]{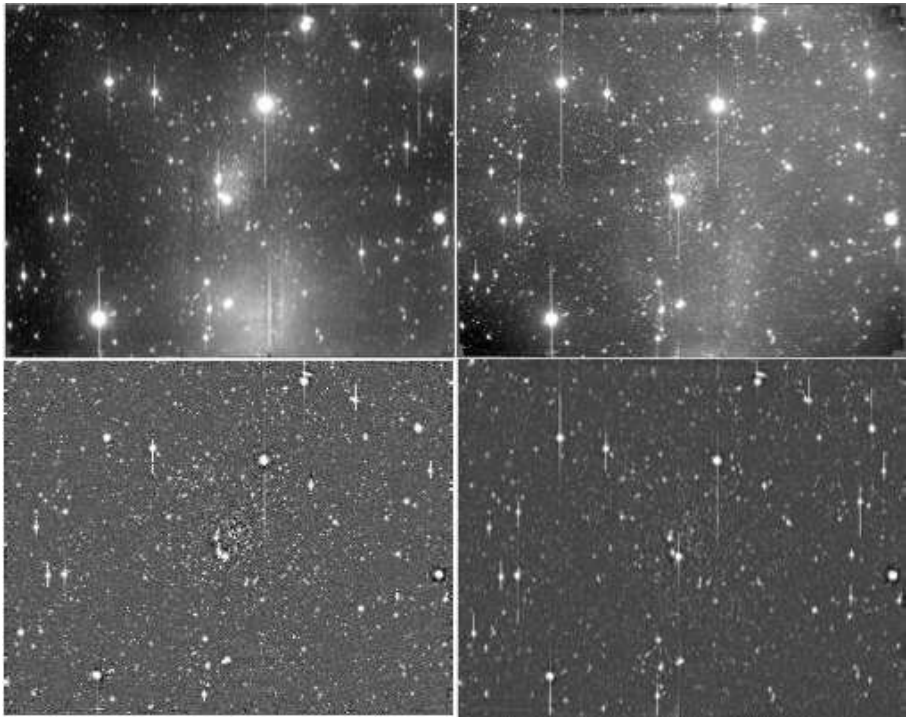}
    \caption{Top panels: the $V-$ (left) and $I_c-$band (right)
    Suprime-Cam fields centered on Andromeda~II ($34$\,arcmins $\times
    27$\,arcmins). North is to the top, and east is to the left. The
    fields are affected by the presence of scattered light from bright
    stars either in, or just off, the field of view. The scattered
    light patterns are complex, but are generally smoothly
    varying. Bottom panels: same as top, where the scattered light has
    been removed using a non-linear equivalent of unsharp masking
    prior to stacking. Low level artifacts and negative halos around
    bright stars still remain, but this has a negligible effect of the
    photometry of the vast majority of objects.}
    \label{ds9}
  \end{center}
\end{figure*}

\begin{figure}
  \begin{center}
    \includegraphics[angle=270, width=8.cm]{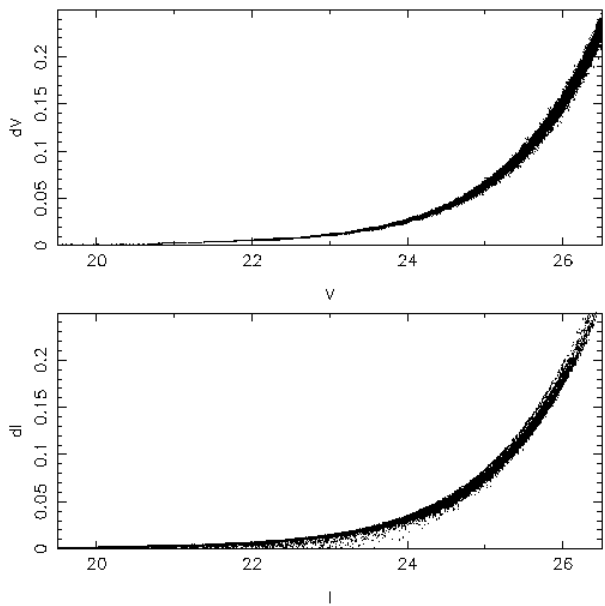}
    \caption{Photometric errors as a function of magnitude for the $V$
    (top panel) and $I$ (bottom panel) data. The data start to become
    incomplete at a signal-to noise ratio of 10, corresponding to $V =
    I \simeq 25.5$.}
    \label{err}
  \end{center}
\end{figure}

During the nights of $3^{{\rm rd}}$ -- $5^{{\rm th}}$ August 2005, we
obtained Johnson -- Cousins $V$ and $I_c-$band imaging of the M31 dSph
galaxies Andromeda~I, II, III, V, VI and VII using Subaru Suprime-Cam
(P.I. N. Arimoto).  Conditions were uniformly excellent, being
photometric throughout and with typical seeing of $0.5$\,arcsecs. At
the start of the night, we also obtained some multi-colour imaging of
the distant transition dwarf galaxy DDO210 and the isolated dSph in
Cetus.  The observations and results for DDO210 are presented in
\cite{mcconnachie2006d}.
 
For And~II, we exposed for a total of $2200$\,seconds in $V$ and
4800\,seconds in $I_c$, split as $5 \times 440$\,seconds and $20
\times 240$\,seconds dithered sub-exposures respectively.  The
telescope was typically offset $\sim 20$\,arcsecs between
sub-exposures. The exposure times were designed to reach below the
horizontal branch to an equivalent depth in the $V$ and $I$ bands. The
final stacked images have sub-arcsec seeing over the whole array,
averaging $0.51$\,arcsec for the $I_c$-band and $0.62$\,arcsec for the
V-band. During stacking, the single frame object catalogues were used
to improve the positional match with respect to the chosen reference
image catalogue by applying an additional $6$ constant linear solution
after the application of the differential WCS. This improved the WCS
solution to approximately one tenth of a pixel precision.

Data were processed using a general purpose pipeline for processing
wide-field optical CCD data (\citealt{irwin2001}). Images were
debiased and trimmed, and then flatfielded and gain-corrected to a
common internal system using clipped median stacks of nightly twilight
flats.  In addition, the $I_c-$band images, which suffer from an
additive fringing component, were also corrected using a fringe frame
computed from the entire series of $I_c$-band exposures taken during
the 3 nights. The top panels of Figure~\ref{ds9} shows the full
$34$\,arcmins $\times 27$\,arcmins field for our reduced $V$ and
$I_c-$band Suprime-Cam images of And~II. North is at the top, and east
is to the left.

The top panels of Figure~\ref{ds9} reveal the presence of scattered
light which affects the final stacked images in an unusual way.  The
light appears to be scattered from bright stars either in, or just
off, the field-of-view.  This leads to complex, but generally smoothly
varying, background light patterns.  However, when the component
images are stacked in a conventional way the induced scattered light
patterns can become disjoint, even after allowing for overall changes
in the background during the stacking process.  To compensate for this
we were forced to remove the smoothly varying background components
prior to stacking using a non-linear equivalent of unsharp
masking. Although this worked well at removing the disjoint patterns,
some of the more rapidly varying spatial components are still visible
as low level artifacts (eg.  the horizontal striations below the
center).  The other drawback of unsharp masking is the low level
negative halo induced around bright stars; however, this has a
negligible effect on the photometry of the overwhelming majority of
individual objects. The bottom two panels of Figure~\ref{ds9} show the
$V$ and $I_c-$band Suprime-Cam images after this correction has been
applied.

For each image frame an object catalogue was generated using the
object detection and parameterisation procedure discussed in
\cite{irwin2004}.  Astrometric calibration of the individual frames
was based on a simple Zenithal polynomial model derived from linear
fits between catalogue pixel-based coordinates and standard
astrometric stars derived from on-line APM plate catalogues.  The
astrometric solution was then used to register the frames prior to
creating a deep stacked image in each passband.  Object catalogues
were then created from these stacked images and objects were
morphologically classified as stellar or non-stellar (or noise-like).
The detected objects in each passband were then merged by positional
coincidence (within $1$\,arcsec) to form a $V, I_c$ combined
catalogue.

We cross-correlated the Suprime-Cam photometry with our earlier
multi-colour INT~WFC photometry of And~II, for which we know the
colour transformations into the Landolt
system\footnote{http://www.ast.cam.ac.uk/~wfcsur/technical/photom/colours}. This
ensures our new photometry is on the same systems as our previous
photometry. By only considering those objects reliably identified as
stellar in all four sets of observations, we find

\begin{eqnarray}
V & = & V^\prime + 0.030 \left(V - I\right)\nonumber\\
I & = & I_c - 0.088 \left(V - I\right)~,
\label{sub2lan}
\end{eqnarray}

\noindent where we now use $V^\prime$ to denote the original Subaru
$V$ filter. These transformations are identical to those derived using
our data for DDO210 (\citealt{mcconnachie2006d}). Figure~\ref{err}
shows our photometric errors as a function of magnitude for our
$V$-band (top panel) and $I$-band (bottom panel) data. Our errors are
$< 0.02$\,mags for $V < 23.6$ and $I < 23.4$. The data start to become
incomplete at a signal-to-noise of 10, corresponding to $V = I \simeq
25.5$.

\section{Stellar Populations and Structure}

\subsection{Colour-magnitude diagrams}

\begin{figure*}
  \begin{center}
    \includegraphics[angle=270, width=14.5cm]{figure3a.ps}
    \includegraphics[angle=270, width=14.5cm]{figure3b.ps}
    \caption{Top panels: Extinction-corrected colour magnitude
    diagrams (left panels) and Hess diagrams (right panels) for
    Andromeda II. Error bars show the average $1\,\sigma$
    uncertainties in the photometry at each magnitude level. The
    dashed boxes in the $V$-band Hess diagram show the colour --
    magnitude cuts used to define the loci of various stellar
    populations. Star symbols show the $(<V>, <(V-I)>)$ locus of
    RR~Lyrae stars in Andromeda~II identified in Da Costa et
    al. (2000) and Pritzl et al. (2004).}
    \label{cmds}
  \end{center}
\end{figure*}

\begin{figure}
  \begin{center}
    \includegraphics[angle=270, width=8cm]{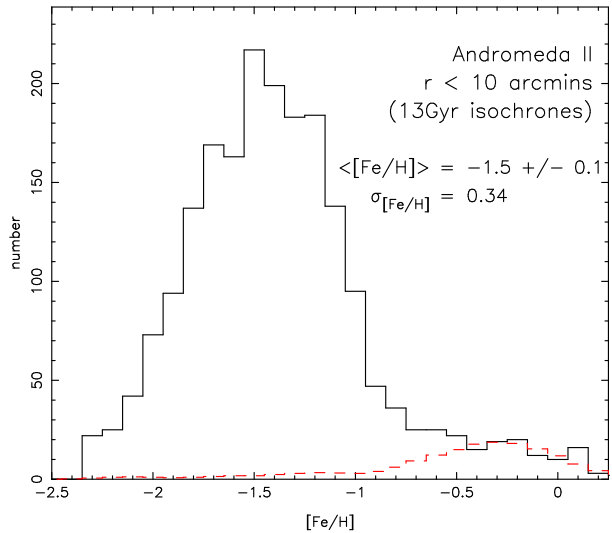}
    \caption{The metallicity distribution function (MDF) for stars in
    the top two magnitudes of the RGB with $r < 10$\,arcmins in
    Andromeda~II (solid histogram). The dashed histogram shows the MDF
    derived for the foreground population ($r > 14$\,arcmins), scaled
    by area. These were created by interpolating between 13\,Gyr
    isochrones from VandenBerg et al. (2006) with $BVRI$
    colour-$T_{eff}$ relations as described by VandenBerg \& Clem
    (2003). The (foreground corrected) mean metallicity is [Fe/H] $=
    -1.5 \pm 0.1$ with $\sigma_{\rm [Fe/H]} = 0.34$\,dex. }
    \label{and2mdf}
  \end{center}
\end{figure}

\begin{figure}
  \begin{center}
    \includegraphics[angle=270, width=8.cm]{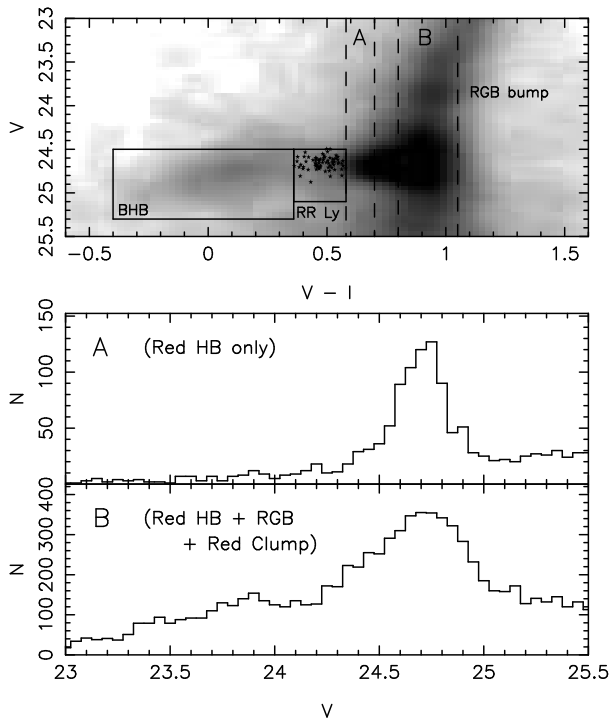}
    \caption{Top panel: an enlargement of the $V$-band Hess diagram of
    Andromeda~II in the region of the horizontal branch (HB). Star
    symbols show the $(<V>, <(V-I)>)$ locus of RR~Lyrae stars in
    Andromeda~II identified in Da Costa et al. (2000) and Pritzl et
    al. (2004). The red HB begins immediately to the right of the
    RR~Lyrae locus. However, the morphology of this feature appears to
    change significantly as a function of colour. The middle and
    bottom panels show luminosity functions in the two strips
    indicated: strip A will only sample red HB stars, whereas strip B
    will additionally sample some of the RGB and a red clump, if
    present. The morphology of the peaks in the two luminosity
    functions are very different, implying either that the spread in
    red HB luminosity becomes much larger to the red, or that an
    additional stellar population (most likely a red clump) is present
    in B but not in A. Note that a RGB bump is also visible in these
    panels at $V \simeq 23.9$. }
    \label{hbpanel}
  \end{center}
\end{figure}

The left panels of Figure~\ref{cmds} show the extinction-corrected
CMDs for And~II, where each star has had its $V$ and $I$ magnitude
corrected for extinction by cross-correlating the position of the star
with the maps of \cite{schlegel1998}. The right panels show the
corresponding Hess diagrams with square-root scaling. Also marked on
the CMDs are error bars which represent the $1\,\sigma$ photometric
uncertainties in our data. These data go several magnitudes deeper
than our earlier INT~WFC imaging in both filters, to a depth
comparable to the earlier HST-WFPC2 imaging of
\cite{dacosta2000}. Suprime-Cam has a $\sim 100$ larger field of view
than HST-WFPC2; $> 35 000$ stars are shown in the CMDs in
Figure~\ref{cmds}, compared to the $\sim 2000$ shown in Figure~5 of
\cite{dacosta2000}. A MW foreground sequence is visible in
Figure~\ref{cmds} at bright magnitudes, starting at $\left(V -
I\right) \simeq 0.6$\,mags. All the other features in the CMD are
produced by stellar populations intrinsic to And~II. The lack of any
obvious bright, main-sequence stars or blue-loop stars suggests that
there has not been any recent (last few Gyrs) star formation. Instead,
all of the prominent populations (the RGB and core helium burning
stars) are signs of intermediate ($2 - 10$\,Gyrs) and/or old ($>
10$\,Gyrs) ages.

\subsubsection{The red giant branch}

In the top left panel of Figure~3 we have divided the broad RGB of
And~II into a `blue' and `red' component.  In general, blue RGB stars
are likely younger and/or more metal-poor than redder RGB stars,
although the unknown effects of the well known age-metallicity
degeneracy leads to difficulty in the interpretation of RGB colour.
We also note the presence of an overdensity of stars on the
RGB at $V \simeq 23.9$, which is most likely due to a RGB bump.

\cite{dacosta2000} showed that And~II has the broadest colour
dispersion in its RGB of any of the Local Group dSph galaxies,
implying $\sigma_{{\rm [Fe/H]}} \sim 0.36$\,dex for a single, co-eval,
ancient population. \cite{cote1999b} spectroscopically measured a
scatter in the metallicity of 50 RGB stars in And~II of $\sigma_{{\rm
[Fe/H]}} = 0.34 \pm 0.1$\,dex, virtually identical to the photometric
measurement. Figure~\ref{and2mdf} shows a metallicity distribution
function (MDF) for $r < 10$\,arcmins derived from our data, where we
have calculated the metallicity of each star in the top two magnitudes
of the RGB by interpolating its position in colour-magnitude space
between $13$\,Gyr isochrones with a range of metallicities (assuming
zero $\alpha-$enhancement). Our technique is standard, and uses the
Victoria-Regina set of isochrones from \cite{vandenberg2006} with
$BVRI$ colour-$T_{eff}$ relations as described by
\cite{vandenberg2003}. The dashed histogram shows the MDF for RGB
stars at $r > 14$\,arcmins, scaled by area, which will be dominated by
foreground stars and which we use as a reference field. Since these
stars are predominantly foreground dwarfs in the Milky Way halo, the
metallicities which we calculate are physically meaningless, and
useful only as a comparison to the main MDF.

The mean metallicity of RGB stars under our age assumption is [Fe/H]
$= -1.5 \pm 0.1$. The dispersion in metallicity is $\sigma_{{\rm
[Fe/H]}} = 0.35 \pm 0.1$\,dex, in excellent agreement with
\cite{dacosta2000} and \cite{cote1999b}, demonstrating good
consistency with these earlier studies. We note, however, that the
mean metallicity and dispersion are sensitive to our age assumption,
reflecting the age-metallicity degeneracy in the colour of the RGB
discussed above. If the stellar populations of And~II are younger than
13\,Gyrs, or if a range in age is present, then the metallicity
estimates we derive will be misleading. Later, we show that this is
nearly certainly the case for And~II.

\subsubsection{The core helium burning stars}

Horizontal branch stars (low mass, core helium burning) are only
present in old stellar populations. The presence of blue HB stars and
RR~Lyrae variables in particular are an unambiguous sign of the
presence of a population which is at least as old as the MW globular
clusters ($> 10$\,Gyrs). The CMDs in Figure~3 clearly show the
presence of a HB at $V \sim 24.7$ which extends far to the blue. The
boundary between blue and red HB stars is marked by the instability
strip: at the HB level, this is marked by the mean colours of RR~Lyrae
variable stars.

The Subaru Suprime-Cam data is unsuitable for the identification of
RR~Lyraes. However, \cite{dacosta2000} identify over 70 RR~Lyrae
variables in their HST-WFPC2 dataset, which are analysed in detail in
\cite{pritzl2004}, and who give the $<V>$ and $<(B - V)>$ magnitudes
for these stars. To compare the position of this RR~Lyrae locus with
our $V,I$ data requires the calculation of the colour transformation
between $(B - V)$ for the HST-WFPC2 data and $(V - I)$ for the
Suprime-Cam data. To this end, we have cross-correlated the HST-WFPC2
dataset with the Suprime-Cam data and identified common objects by
positional coincidence within 1\,arcsec. We apply the additional
constraint that the photometric uncertainties in the ground-based data
should be less than $0.05$\,magnitudes in each filter. We then
determine the relation between $(B - V)$ and $(V - I)$, and find it is
well fit by the linear relation

\begin{equation}
\left(V - I\right) = 0.837 \left(B - V\right) + 0.175~.
\end{equation}

\noindent The star symbols in the top right panel of Figure~3 show the
$\left(<V>, <(V - I)>\right)$ locus of the RR~Lyrae stars identified
in the earlier HST-WFPC2 study. The rectangle surrounding these points
shows the colour-magnitude cuts we use to approximate the position of
this locus. The large rectangle immediately to the blue of this
defines the colour-magnitude cuts used to isolate the blue
HB. Likewise, the red HB will lie immediately to the red of the
RR~Lyrae locus. However, the structure of the CMD in this region is
quite complex.

The top panel of Figure~\ref{hbpanel} shows an enlarged version of the
$V-$band Hess diagram at the level of the HB. The RR~Lyrae and blue HB
are marked as before. To the red of the RR~Lyrae locus, a significant
overdensity of stars is present. Some of these will clearly be red HB
stars; however, the width of this overdensity increases significantly
towards the red. To illustrate this, in the middle panel we have
plotted the luminosity function of stars in a strip in colour space
immediately redder than the RR~Lyrae locus (A) and compared this to
the luminosity function of stars in another strip which is even redder
(B) in the bottom panel. Strip A is likely to only sample red HB
stars; however, strip B additionally sample a significant number of
RGB stars, and will also sample red clump stars, if present. This
latter stellar population are also core helium burning stars, but they
have a significantly higher mass than their low mass, HB
counterparts. Their presence would therefore imply the existence of an
intermediate-age stellar population (see \citealt{girardi2001} for a
comprehensive theoretical study of the red clump).

The peak of the luminosity functions in Figure~\ref{hbpanel}
correspond to core helium burning stars in And~II; however, the peaks
clearly have vastly different dispersions and shapes. Therefore,
either the spread in the luminosity of red HB stars varies
dramatically as a function of colour, or the broader peak in B is due
to the presence of an additional stellar population with similar
luminosities and colours to the HB. We are unaware of a physical
mechanism which could cause the former to occur. On the other hand,
\cite{dacosta2000} independently implied that And~II had to possess an
intermediate-age population (of age $6 - 9$\,Gyrs) in order to explain
the overall HB morphology of And~II as revealed by their HST-WFPC2
data. Such a population would be expected to possess a red
clump. Therefore, we conclude that the morphology of the region of the
CMD shown in Figure~\ref{hbpanel} is best and most naturally explained
by the presence of a red clump population in this galaxy. This is the
first detection of a red clump in any of the M31 dSphs, which are often
characterised as having no significant intermediate age
populations. Colour-magnitude cuts to define the red HB locus and red
clump locus are shown in the top-right panel of Figure~3 (the red
clump locus will also contain significant contributions from red HB
and RGB stars). We will examine the luminosity and colours of all the
core helium burning stars in more detail in Section~3.5.

\subsection{Radial gradients}

\begin{figure}
  \begin{center}
    \includegraphics[angle=270, width=8cm]{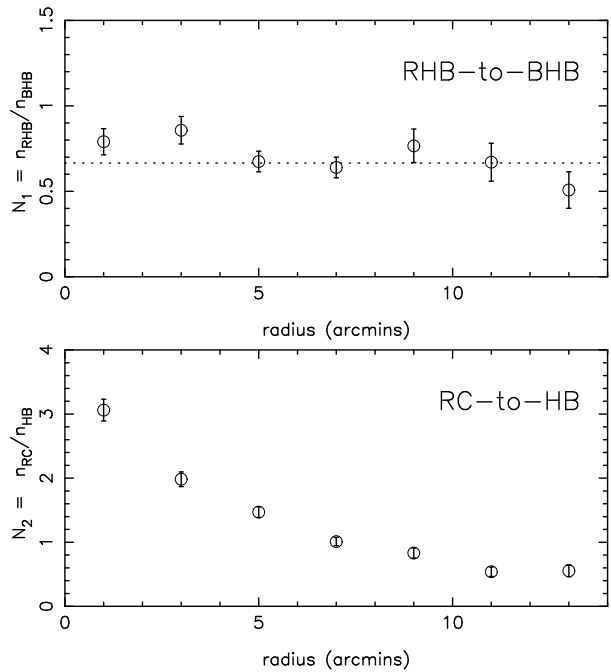}
    \caption{Top panel: the ratio of the number of red HB stars to the
    number of blue HB stars, as a function of radius in
    Andromeda~II. The dotted line is the mean value of this ratio
    averaged over all radii.  Clearly, this ratio does not change
    significantly over a large radial range, implying that the blue HB
    and red HB stars posses similar radial distributions. Bottom
    panel: the ratio of the number of red clump stars to the number of
    (blue + red) HB stars.  The ratio decreases significantly as a
    function of radius, demonstrating that these populations possess
    different spatial distributions, such that the red clump stars are
    significantly more concentrated than the HB stars.}
    \label{hbrad}
  \end{center}
\end{figure}

\subsubsection{The red-to-blue horizontal branch profile}

\cite{dacosta2000} and \cite{harbeck2001} looked for population
gradients in And~II by examining the change in the number of blue HB
stars to the total number of (blue + red) HB stars over the radial
extent of the HST-WFPC2 field. Both papers concluded that no
significant radial variation in this quantity was present in their
data.

Using our colour cuts to define blue and red HB stars, we have
calculated an analogous quantity to \cite{dacosta2000}, $N_1 =
n_{RHB}/n_{BHB}$ (the ratio of the number of red HB stars to the
number of blue HB stars) as a function of radius. The radial extent of
And~II probed by our data is significantly larger than that probed by
the HST-WFPC2 data, and so should display any large-scale variations
in this ratio if they are present. We count stars in annuli of width
1\,arcmin centered on And~II, and display our results in the top panel
of Figure~\ref{hbrad}. $N_1$ remains approximately flat, with no
significant deviation from the mean value, out to 13\,arcmins from the
center of And~II. This implies that all of the obviously old stellar
populations in And~II have very similar radial distributions, and is
in agreement with the earlier studies.

\subsubsection{The red clump to horizontal branch profile}

We now define an analogous quantity to $N_1$, $N_2 = n_{RC}/n_{HB}$
(the ratio of the number of red clump stars to the number of all HB
stars). This will probe the relative contribution of low-to-high mass
helium burning stars in And~II as a function of radius.

The profile of $N_2$ is shown in the bottom panel of
Figure~\ref{hbrad}. In contrast to the ratio of red to blue HB
stars, the ratio of red clump to HB stars declines significantly and
continually as a function of radius by nearly a factor of 6 over a
13\,arcmin radial range. Clearly, the low and high mass helium burning
stars in And~II have very different radial distributions, such that
the red clump stars are more centrally concentrated than the HB
stars. This then implies that the intermediate age stellar populations
in And~II are significantly more centrally concentrated than the old
stellar populations. And~II clearly possesses a very strong radial
gradient in its stellar populations.

\subsection{Spatial distributions}

\begin{figure*}
  \begin{center}
    \includegraphics[angle=0, width=14.cm]{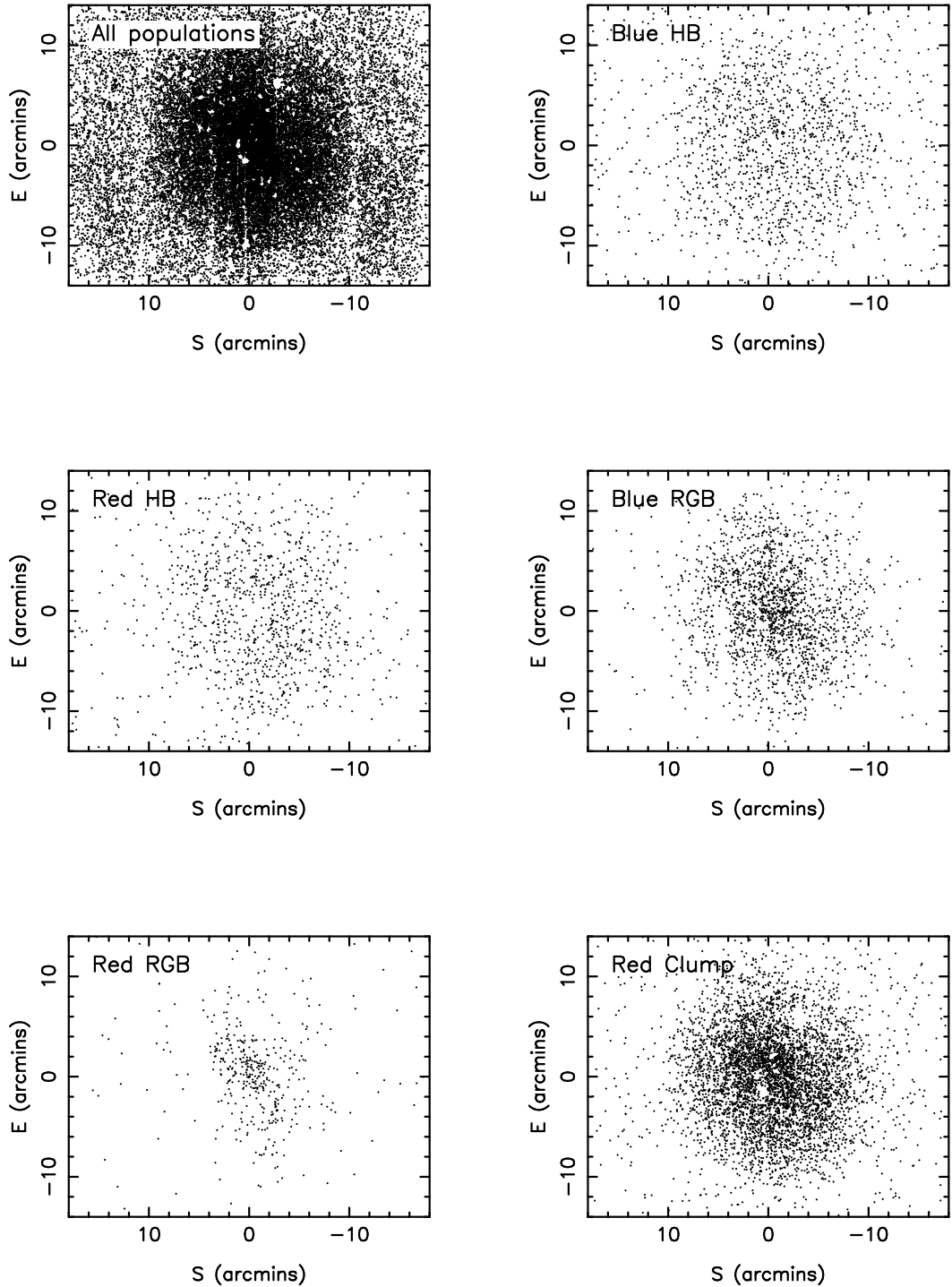}
    \caption{The spatial distribution of stellar populations in
    Andromeda~II. The top right panel consists of all stars brighter
    than our $\sim 90\,\%$ completeness limits. The five remaining panels
    show the distribution of various stellar populations, defined by
    the colour-magnitude loci shown in Figure~3. The blue and red HB
    populations both posses diffuse and extended distributions. In
    contrast, the red RGB stars have a very different spatial
    distribution which is significantly more concentrated and less
    extended than the HB distribution. The blue RGB and red clump
    distributions are relatively similar to each other, but are
    different to the other sub-populations. }
    \label{maps}
  \end{center}
\end{figure*}

Figure~\ref{maps} shows maps of the spatial distribution of stellar
sources satisfying various colour-magnitude cuts in And~II, projected
into the tangent plane of this galaxy.  The top panels of
Figure~\ref{maps} show the spatial distribution of all stars in And~II
brighter than our $\sim 90\,\%$ incompleteness limits. And~II is a very
large, extended spheroid of stars with a slight ellipticity of
$\epsilon \sim 0.2$ (MI6). Foreground stars will dominant in the outer
parts of the Suprime-Cam field, but the tidal extent of And~II is
sufficiently large ($r_t = 19.8$\,arcmins; MI6) that some of its stars
will be present at all radii probed.

The remaining five panels in Figure~\ref{maps} correspond to stars in
the stellar loci indicated in Figure~3. Given the tight
colour-magnitude cuts used to isolate each population, the
contamination from foreground stars in these panels is generally
small. Some of the irregularity of the stellar distributions can be
traced back to the presence of saturated stars in the field, which can
cause holes to appear in the spatial distributions; for example, near
the center of the red and blue HB distributions.

The red and blue HB both have extended, diffuse stellar distributions
which do not obviously increase in density at small radius, as is
common for spheroidal profiles. In comparison, the red clump and blue
RGB populations are approximately as extended as the HB distributions,
but are much more centrally concentrated. This naturally explains the
gradient observed in Section~3.2.2. The red RGB, on the other hand, is
far less extended than all the other stellar distributions, and is
concentrated in the central regions of And~II. The contrast between
the HB distribution and the red RGB distribution is striking.

\subsection{Radial profiles}

\begin{figure*}
  \begin{center}
    \includegraphics[angle=270, width=16.cm]{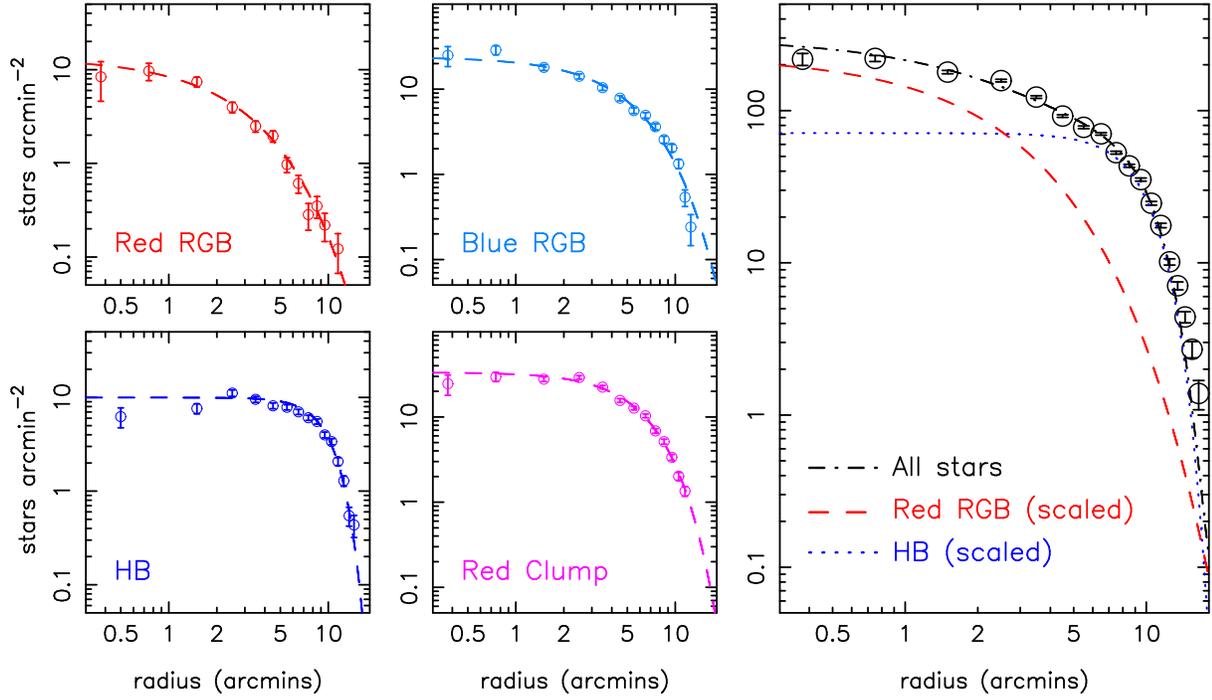}
    \caption{Left and center panels: radial distributions in
    elliptical annuli of the stellar populations in Andromeda~II
    defined by the colour-magnitude loci in Figure~3 (the blue HB,
    RR~Lyrae and red HB have been grouped together). Best-fitting
    Sersic profiles have been overlaid as dashed lines. The red RGB
    and HB profiles are most distinct; the former is nearly
    exponential whereas the latter is approximately constant density
    out to $\sim 10$\,arcmins. Right panel: the overall radial profile
    of Andromeda~II derived using all stars brighter than our $\sim
    90\,\%$ completeness limit. Remarkably, the overall profile of
    this galaxy is very well described by an appropriately weighted
    addition of the HB and red RGB profiles. The fit obtained on doing
    this is shown as a dot-dashed line, whereas the scaled HB and red
    RGB profiles are shown as dotted and dashed lines,
    respectively. This suggests that Andromeda~II is composed of two
    dominant components; a centrally concentrated, nearly exponential
    component best traced by red RGB stars, and an extended, nearly
    constant density component best traced by the HB stars.}
    \label{radial}
  \end{center}
\end{figure*}

We robustly quantify the spatial distribution of the stellar
populations shown in Figure~\ref{maps} by constructing radial profiles
for each population. We count stars in elliptical annuli with the
average ellipticity and position angle of And~II ($\epsilon = 0.2,
\theta = 34^\circ$; MI6). While the maps in Figure~\ref{maps} show
that these quantities change for And~II depending upon the stellar
population examined, adopting annuli of fixed shape and orientation
for all the populations allows for a more robust and meaningful
comparison between the resulting profiles.

The radial profiles for the individual stellar populations are shown
in the left and center panels of Figure~\ref{radial} in log-log
space. The HB profile consists of the blue HB, RR~Lyrae and red HB
stellar loci, since these have been shown to possess similar
distributions. The radial profile for all stars above the $\sim 90\,\%$
completeness limit of our data is shown in the right panel of
Figure~\ref{radial}.

A foreground correction has been applied to the overall profile by
comparison with the (foreground-corrected) radial profile in MI6. We
have assumed that the core and tidal radii of the best-fitting King
model derived by MI6 should also fit the outer regions of our
Suprime-Cam data. Our free parameters in this fit are a vertical scale
factor and a constant representing the foreground contamination in the
Suprime-Cam data. The best-fit value of the latter can then be
subtracted from the Suprime-Cam profile. Although no foreground
correction has been applied to the profiles for the individual
populations, the tight colour-magnitude cuts which define these
populations ensure that the foreground contamination is minimal.

Since the radial profiles in the left and center panels of
Figure~\ref{radial} show significant differences, we choose the
flexible Sersic profile to parameterise them all. The projected
density of this distribution function is given by

\begin{equation}
I\left(r\right) = I_0 \exp\left[-\left(r/r_o\right)^{1/n}\right]~.
\end{equation}

\noindent $n$ is a shape parameter ($n = 1$ corresponds to an
exponential profile and $n = 4$ corresponds to a de Vaucouleurs
profile) and $r_o$ is a scale-radius. $I_0$ is a normalisation
constant and is equivalent to the central surface brightness. The
best-fit Sersic profiles to the radial profiles of the individual
stellar populations are shown as dashed lines in the left and center
panels of Figure~\ref{radial}. For each population, we find the
following best-fit parameters:

\begin{itemize}
\item{Red clump: $r_o = 5.76 \pm 0.84$, $n = 0.60 \pm 0.02$}
\item{Blue RGB: $r_o = 4.34 \pm  0.47$, $n = 0.79 \pm 0.02$}
\item{Red RGB: $r_o = 2.18 \pm 0.78$, $n = 1.03 \pm 0.29$}
\item{HB: $r_o = 10.00 \pm 8.48$, $n = 0.30 \pm 0.11$}
\end{itemize}

The red clump and blue RGB are the most similar out of these 4
profiles, although even these are statistically different from each
other. The red clump profile has a slightly more steeply declining
profile than the blue RGB, and a slightly larger scale length. Both
these profiles are very distinct from the red RGB profile, which has
an essentially exponential ($n = 1$) profile with a scale length of
only $2.18 \pm 0.78$ arcmins ($417 \pm 152$\,pc). This is much more
similar to the typical overall radial profile of dSph galaxies in the
Local Group (\citealt{irwin1995}, MI6). We note with interest that the
exponential scale length of this {\it population} is a factor of $\sim
2$ larger than the exponential scale length of the {\it overall}
radial profiles of most of the MW dSph population
(\citealt{irwin1995}).

The radial profile of the HB component is very unusual. The slight
decline in the stellar density of this component at small radii is an
artifact of the few saturated stars in this region which produce holes
in the spatial maps in Figure~\ref{maps}.  The scale radius of the HB
profile is not well constrained, due to the unusual, nearly constant
density shape of the profile. The value of the power-law index, $n =
0.30 \pm 0.11$ produces a very steeply declining profile at large
radius. Thus, the HB population has an approximately constant density
profile out to large radius before it essentially truncates.

Although the radial profiles of the red RGB and HB can both be
described by Sersic profiles, the two bear no similarity to each other
at all, as even a cursory glance of Figure~8 makes clear. A K-S test on
the red RGB and HB cumulative radial profiles shows that the two are
absolutely inconsistent with being drawn from the same underlying
distribution at extremely high significance ($>>99.99\,\%$).

Given that the red RGB has an exponential profile, and the HB has a
constant density profile, it is very likely that the two must have
very different dynamical properties. For example, if they are both in
equilibrium with the overall potential of the dSph, then they cannot
both have the same kinematic properties (velocity dispersion profile,
orbital anisotropy etc) and have such vastly different density
profiles. This is because they will not both satisfy Jeans equation,
which requires a balance between the overall potential, the density
distribution of the tracer population and the kinematic properties of
the tracer population. On the other hand, if one or both of the
populations is not in equilibrium, then there is no reason to expect
that they will have similar kinematics in the first place. It is very
likely therefore that the red RGB population and the HB population
belong to two kinematically distinct components in And~II.

Compelling evidence for this interpretation comes from the overall
radial profile displayed in the right panel of Figure~\ref{radial}. The
dot-dashed line is the best-fit to the data obtained on fitting a
weighted sum of the HB and red RGB profiles, that is the function

\begin{equation}
F(r) = a_1 I_{HB}(r) + a_2 I_{RRGB}(r)
\end{equation}

\noindent where $I_{HB}(r)$ and $I_{RRGB}(r)$ are the best fitting
Sersic profiles to the HB and red RGB populations found previously,
and $a_1$ and $a_2$ are scaling constants found through a
least-squares fit to the overall profile. If, as we postulate, the HB
and red RGB populations trace two distinct components in And~II, then
the sum of the radial profiles of these components must equal the
overall radial profile of And~II.

The dot-dashed line in the right panel of Figure~\ref{radial} is a
remarkably good fit to the overall profile of And~II. The only slight
deviation of this line from the data is at large radius, where we are
particularly sensitive to the foreground correction we applied (which
was not measured directly from our data). Therefore, this deviation
seems small as it is only of order $1$\,star\,arcmin$^{-2}$, and the
fit is good over 2.5\,dex in stellar density. 

The individual, weighted Sersic profiles of the HB and red RGB
populations, $a_1 I_{HB}(r)$ and $a_2 I_{RRGB}(r)$ respectively, are
shown in the right panel of Figure~\ref{radial} as dotted and dashed
lines respectively. Inside of 2 -- 3\,arcmins, the exponential
component traced by the red RGB stars dominates, whereas outside of
this radius the extended component traced by HB stars begins to
dominate. This naturally explains the peculiar profile for And~II
measured by MI6, where a factor of two excess of stars was seen at $r
\lesssim 2$\,arcmins upon the extrapolation of the best-fitting
density profile at large radius to small radius: at large radius, the
extended component dominates, whereas at small radius an additional
component contributes to the density profile, and dominates over the
extended component. In Section~3.5, we present additional evidence in
the ages and metallicities of the stellar populations of And~II which
supports the presence of two distinct components. We note that the
blue RGB and red clump profiles are also able to be approximated by a
weighted sum of the red RGB and HB profiles, as is required in this
model.

The absolute magnitude of And~II obtained by MI6 upon integrating the
surface brightness profile is $M_V = -12.6 \pm 0.2$. By using the
best-fit values of $a_1$ and $a_2$, it is straightforward to calculate the
expected luminosities of the two postulated components. We find that
the extended component contributes $\sim 74\,\%$ of the light in
And~II, while the exponential component will contribute
$\sim 26\,\%$ of the light. This translates into magnitudes of $M_V
\sim -12.3$ and $-11.2$ for the extended and concentrated components,
respectively. The corresponding central surface brightnesses are
$\mu_{0,V} \sim 26.1$ and $24.8$\,mags\,arcsec$^{-2}$, respectively,
where the total central surface brightness of And~II has been taken to
be $24.5$\,mags\,arcsec$^{-2}$ (MI6).

Table~1 lists the above parameters for And~II, along with error
estimates. The uncertainties in $M_V$ and $L_V$ are obtained by
propagating the uncertainties in the absolute magnitude of And~II,
$a_1$ and $a_2$. They are likely to be underestimates of the true
uncertainties, since the adopted profiles of each component also have
considerable uncertainties associated with them. Nevertheless, these
values illustrate the approximate contribution of each component to
the overall composition of And~II.

\subsection{The ages and metallicities of Andromeda~II}

\begin{figure*}
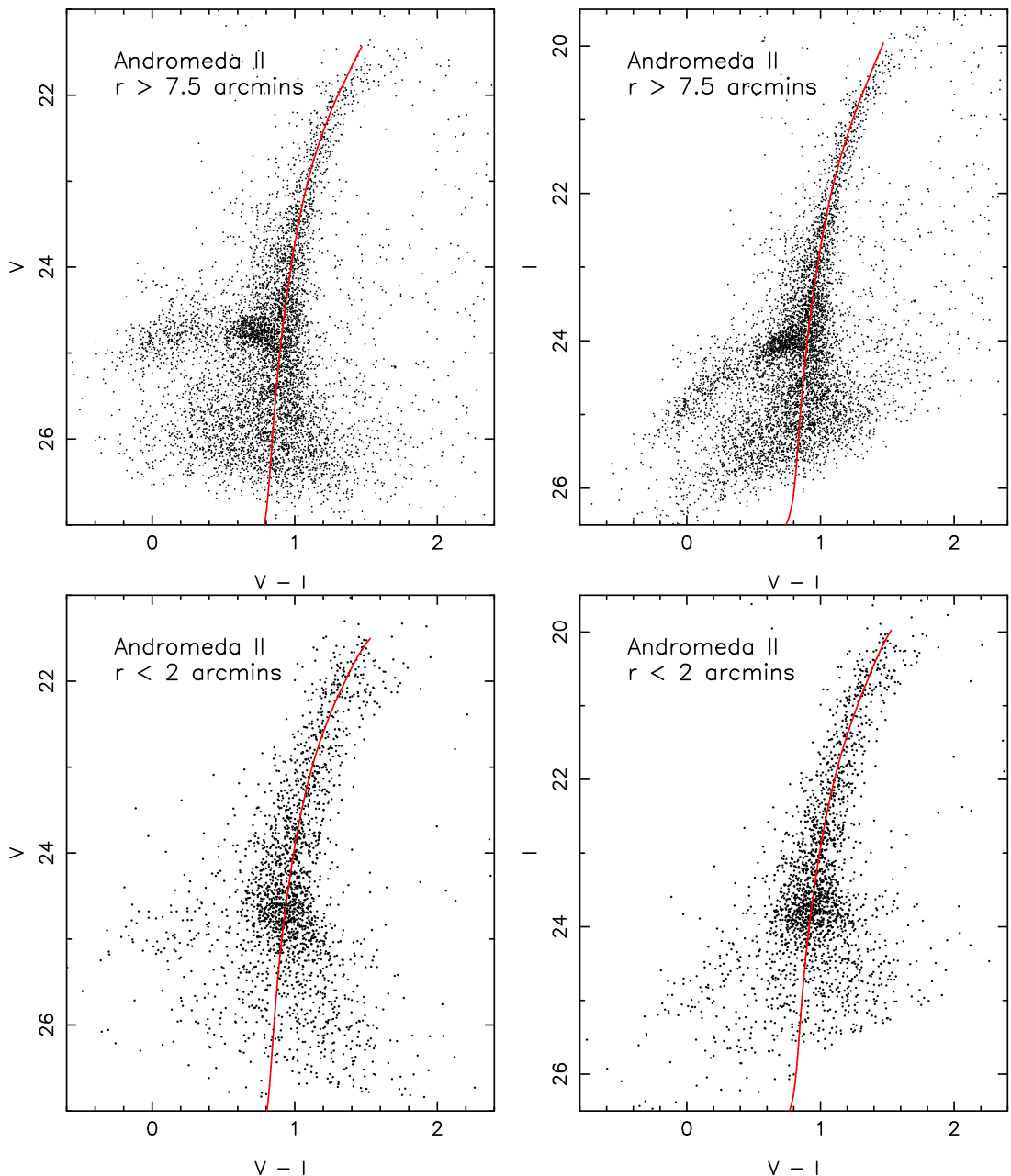

  \begin{center}
    \includegraphics[angle=270, width=14.5cm]{figure9a.ps}
    \includegraphics[angle=270, width=14.5cm]{figure9b.ps}
    \caption{Extinction-corrected $V$ vs $\left(V - I\right)$ and $I$
    vs $\left(V - I\right)$ CMDs for Andromeda~II for $r >
    7.5$\,arcmins (top panels) and $r < 2$\,arcmins (bottom panels).
    The former will predominantly sample the stellar populations of
    the extended component of Andromeda II, whereas the latter will
    predominantly sample the stellar populations of the centrally
    concentrated, exponential component. A 13\,Gyr isochrone with
    [Fe/H] $= -1.5$, from the Victoria - Regina set of isochrones, is
    overlaid to aid visual comparison. The differences between the
    CMDs split in this way are striking, particularly regarding the
    width of the RGB and the morphology of the core helium burning
    stars.}
    \label{split}
  \end{center}
\end{figure*}

\begin{figure*}
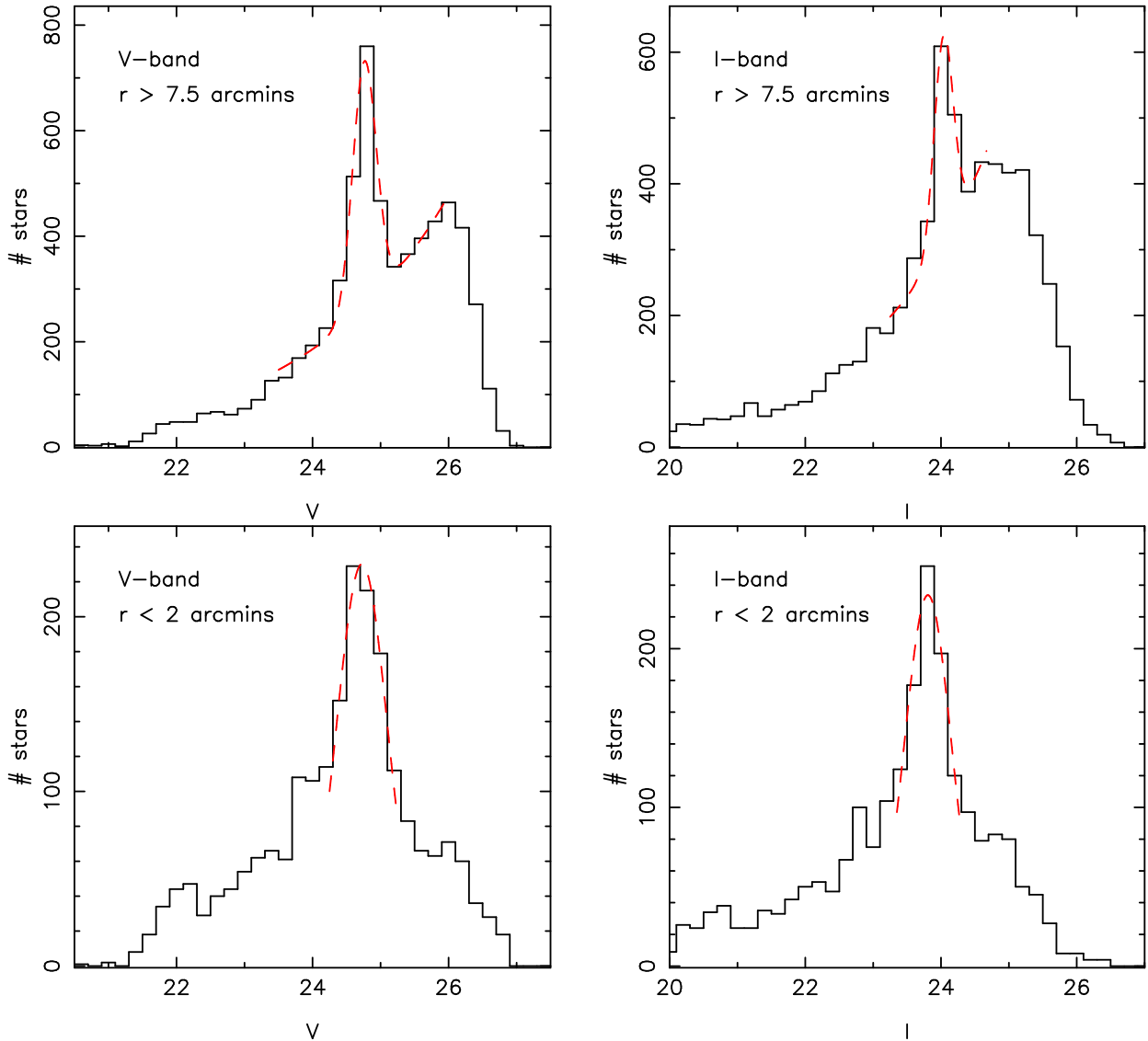

  \begin{center}
    \includegraphics[angle=270, width=16.5cm]{figure10a.ps}
    \includegraphics[angle=270, width=16.5cm]{figure10b.ps}
    \caption{Top panels: Extinction-corrected $V$ and $I$-band
    luminosity functions for stars in Andromeda~II with $r >
    7.5$\,arcmins (which predominantly sample the extended
    component). The dashed lines represent the best-fits of a power
    law and Gaussian to the RGB in the region of the red clump/HB
    level. The Gaussians are centered at $V_{HB} = 24.76 \pm 0.01$ and
    $I_{HB} = 24.04 \pm 0.01$, with dispersions of $\sigma_V = 0.17
    \pm 0.01$ and $\sigma_I = 0.14 \pm 0.01$. Bottom panels:
    Extinction-corrected $V$ and $I$-band luminosity functions for
    stars in Andromeda~II with $r < 2$\,arcmins (which predominantly
    sample the exponential component). The dashed lines represent the
    best-fits of a Gaussian to the red clump/HB level. The Gaussians
    are centered at $V_{HeB} = 24.73 \pm 0.02$ and $I_{HeB} = 23.81
    \pm 0.02$, with dispersions of $\sigma_V = 0.37 \pm 0.05$ and
    $\sigma_I = 0.34 \pm 0.02$.}
    \label{lfs}
  \end{center}
\end{figure*}

\begin{figure*}
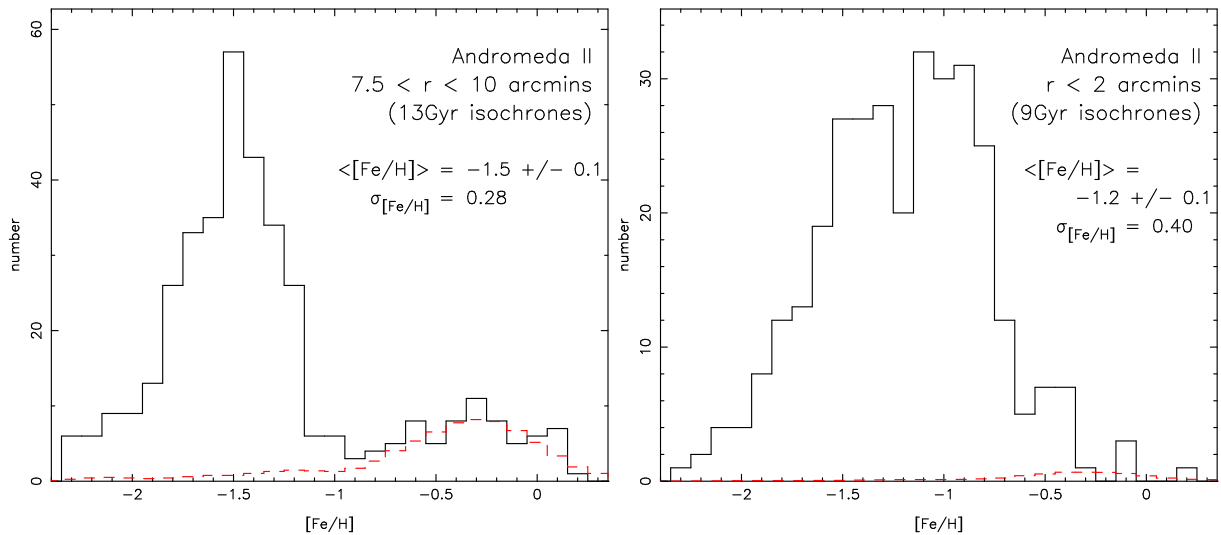

  \begin{center}
    \includegraphics[angle=270, width=8cm]{figure11a.ps}
    \includegraphics[angle=270, width=8cm]{figure11b.ps}
    \caption{Left panel: The metallicity distribution function (MDF)
    for stars in the top two magnitudes of the RGB with $10 > r >
    7.5$\,arcmins in Andromeda~II (solid histogram). This radial range
    predominantly samples stars in the spatially extended component in
    Andromeda~II. This was created by interpolating between 13\,Gyr
    isochrones from VandenBerg et al. (2006) with $BVRI$
    colour-$T_{eff}$ relations as described by VandenBerg \& Clem
    (2003). The mean metallicity is [Fe/H] $= -1.5 \pm 0.1$ with
    $\sigma_{\rm [Fe/H]} = 0.28$\,dex. Right panel: The same, but for
    RGB stars with $r < 2$\,arcmins in Andromeda~II (solid histogram),
    and interpolating between $9$\,Gyr isochrones. This radial range
    predominantly samples stars in the exponential
    component in Andromeda~II. The mean metallicity is [Fe/H] $= -1.2
    \pm 0.1$ with $\sigma_{\rm [Fe/H]} = 0.40$\,dex. The dashed
    histograms in both panels show the relevant MDF derived for the
    foreground population ($r > 14$\,arcmins), scaled by area.}
    \label{mdfs}
  \end{center}
\end{figure*}

\begin{figure*}
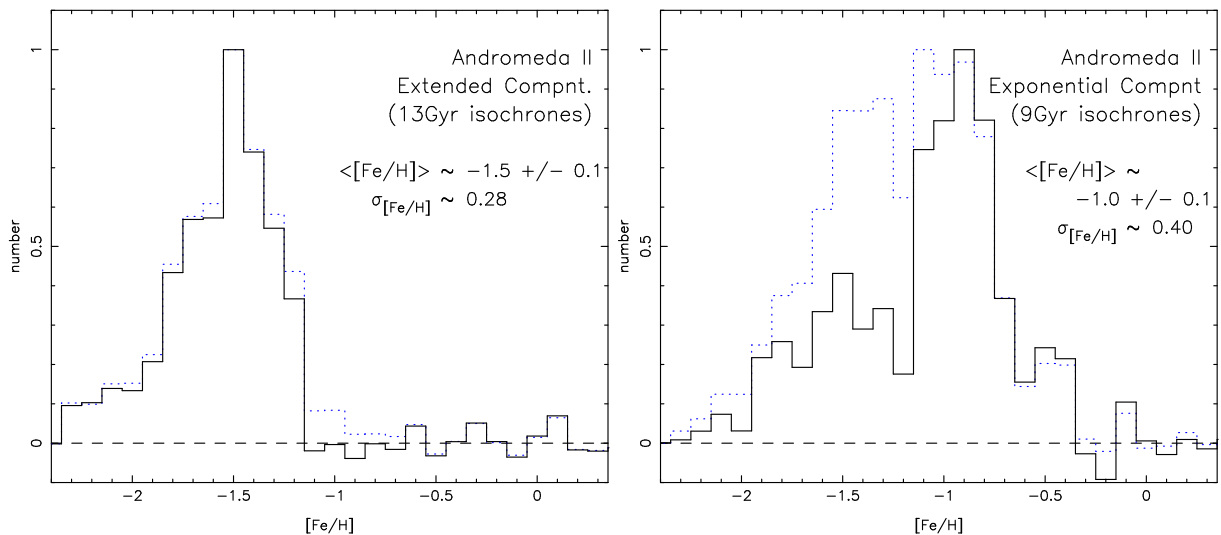

  \begin{center}
    \includegraphics[angle=270, width=8cm]{figure12a.ps}
    \includegraphics[angle=270, width=8cm]{figure12b.ps}
    \caption{MDFs for the spatially extended and exponential
    components in And~II (solid histograms; left and right panels,
    respectively). The dotted histograms show the MDFs in the outer
    and inner regions of And~II for comparison. See text for details.}
    \label{2comps}
  \end{center}
\end{figure*}

The analysis in Section~3.4 suggests that different stellar components
with different characteristic stellar populations dominate And~II at
small and large radius. To investigate this possibility further, we
show in Figure~\ref{split} the extinction corrected CMDs of And~II
split by radius. The top panels show the And~II CMDs for stars with $r
< 2$\,arcmins. According to the fits in Section~3.4, the exponential
component dominates the extended component by a factor of
approximately 2:1 inside this radius. The bottom panels of
Figure~\ref{split} show the And~II CMDs for stars with $r >
7.5$\,arcmins. The extended component is expected to dominate the
concentrated component by a factor of approximately 8:1 over this
radial range. To aid visual comparison, a 13\,Gyr isochrone with
[Fe/H] $= -1.5$ from the Victoria - Regina set of isochrones, shifted
to the distance modulus of And~II, has been overlaid. The difference
in the stellar populations divided by radius is striking, particularly
regarding the width of the RGB and the morphology of the HB/red clump
level. Each of these features contains information regarding the
average ages and metallicities of the dominant stellar populations at
these radii.

\subsubsection{Populations at large radii}

The large majority of stars in the top panels of Figure~\ref{split}
are expected to belong to the spatially extended stellar
component.  A blue HB population occupying the same colour-magnitude
locus as defined earlier is clearly visible in these panels. These
stars unambiguously determine the age of this population to be as old
as the MW globular clusters ($10 - 13$\,Gyrs).

Additionally information on the average age and metallicity come from
the colour and magnitude of the core helium burning
stars. \cite{girardi2001} show that the mean $I$-band magnitude of
core helium burning stars is primarily dependent on age, whereas the
mean $\left(V - I\right)$ colour is primarily dependent on
metallicity. While their study is focused towards understanding the
evolution of red clump stars, the results they provide extend to old
(low mass) helium burning stars as well. The top panels of
Figure~\ref{lfs} show the extinction-corrected $V$- and $I$-band
luminosity functions of all stars in And~II with $r >
7.5$\,arcmins. The peak of stars in both panels is due to the HB. The
dot-dashed lines correspond to the best fits of a Gaussian (which
models the luminosity profile of the HB ) and a power law (which
models the slope of the RGB in this region). The peak magnitude of the
Gaussians are the mean magnitudes of the HB stars. We find $V_{HB} =
24.76 \pm 0.01$ and $I_{HB} = 24.04 \pm 0.01$, with corresponding
dispersions of $\sigma_V = 0.17 \pm 0.01$ and $\sigma_I = 0.14 \pm
0.01$. This implies $M_V^{HB} = 0.69 \pm 0.06$, $M_I^{HB} = -0.03
\pm 0.06$, and $\left(V - I\right)_o = 0.72 \pm 0.02$. The
uncertainties in the magnitudes are dominated by the uncertainty in
the distance modulus of And~II.

The relationships derived by \cite{girardi2001} between the colour,
magnitude, age and metallicity of core helium burning stars are shown
in their Figure~1. By comparing this with our observed values for the
luminosity and colour of the HB stars we find that either these stars
belong to a very young, comparatively metal-rich population ($<
2$\,Gyrs, [Fe/H] $> -0.5$), or they belong to a very old, metal-poor
population ($\sim 13\,$Gyrs, $-1.7 \lesssim [Fe/H] \lesssim
-1.3$). However, the blue HB stars show that this population must be
old, favouring the more metal-poor estimate.

A final constraint can be placed on the metallicity of the stars in
this radial range by constructing a MDF from interpolation of the
position of stars on the RGB. We use exactly the same technique as
before, adopting 13\,Gyr isochrones for the interpolation procedure to
ensure consistency with the HB age information. The result is shown in
the left panel of Figure~\ref{mdfs} (for stars with $7.5 < r <
10$\,arcmins). Also shown in this panel as a dashed histogram is the
MDF for stars with $r > 14$\,arcmins, which is dominated by foreground
stars. Clearly, the influence of foreground contamination on this MDF
is minimal. The implied median metallicity of the (foreground
corrected) MDF is [Fe/H] $= -1.5 \pm 0.1$, with a dispersion of
$\sigma_{\rm [Fe/H]} = 0.28$\,dex. The mean metallicity is the same as
we derived in Section 3.1.1, although the MDF for stars at large radii
is notably narrower than the overall MDF for And~II
(Figure~\ref{and2mdf}). The excellent agreement between the age and
metallicity estimates from the HB and the metallicity estimate of the
RGB demonstrates good consistency in our analysis.

\subsubsection{Populations at small radii}

The bottom panels of Figure~\ref{split} show the CMDs for the inner
few arcminutes of And~II, which we expect is dominated by the
exponential stellar component. However, approximately one-third of the
stars are expected to belong to the spatially extended component. The
RGB in this radial range is extremely broad, and the morphology of the
HB/red clump is very different to any of the CMDs examined
previously. Some stars occupy the locus of the blue HB defined
earlier: we count a total of $56 \pm 8$ such stars. However, our
earlier analysis of the radial profiles shows that the expected number
density of HB stars at this radii is $\sim 10$\,stars\,arcmin$^{-2}$
(Figure~\ref{radial}), or a total of $100$ HB stars in the area probed
by these CMDs. Of these, $\sim 43\,\%$ are expected to be blue HB
stars. We concluded earlier that the HB stars traced an extended
component; the expected contribution of blue HB stars from the
extended component at $r < 2$\,arcmins is therefore $\sim 43 \pm 7$,
which is consistent with the number we actually observe. We cannot
rule out the possibility that the exponential component has a very
weak blue HB, but this cannot contribute more than $\sim 10\,\%$ of
the total number of blue HB stars in And~II.

The lower panels of Figure~\ref{lfs} show the extinction-corrected
luminosity functions for stars with $r < 2$\,arcmins.  The dashed
lines show the best-fit Gaussians to the red clump/HB peaks. The mean
magnitudes of the core helium burning locus are $V_{HeB} = 24.73 \pm
0.02$ and $I_{HeB} = 23.81 \pm 0.02$, with dispersions of $\sigma_V =
0.37 \pm 0.05$ and $\sigma_I = 0.34 \pm 0.02$. These correspond to
$M_V^{HeB} = 0.66 \pm 0.06$, $M_I^{HeB} = -0.26 \pm 0.06$ and
$\left(V - I\right)_o = 0.92 \pm 0.03$. By comparing the $\left(V -
I\right)_o$ colour with Figure~1 of \cite{girardi2001}, we imply a
mean metallicity of [Fe/H] $\sim -0.7$ and a mean stellar age of $\sim
9$\,Gyrs, with a plausible range given the uncertainties of $7 -
10.5$\,Gyrs. The mean $I$-band magnitude of the red clump is notably
brighter than the mean HB luminosity measured previously, implying a
stellar population that is significantly younger by at least 3 or so
Gyrs. This finding is in good agreement with \cite{dacosta2000}, who
implied that there was a younger population in And~II and that its age
was probably in the range $\sim 6 - 9$\,Gyrs.

We derive an MDF for the inner regions in the same way as before,
except we now use isochrones which are 9\,Gyrs old to agree with the
red clump age estimate. The MDF is shown in the right panel of
Figure~\ref{mdfs}, and the dashed histogram shows the corresponding
MDF for the foreground population scaled by area. The median
metallicity is [Fe/H] $= -1.2 \pm 0.1$, with a very broad dispersion
of $\sigma_{{\rm [Fe/H]}} = 0.4$\,dex. Clearly, the broad metallicity
dispersion for And~II measured by \cite{dacosta2000} and
\cite{cote1999b} reflects the large dispersion present in the central
regions of this galaxy, and which is not found at large radius.


\subsubsection{MDFs of the two components}

In our two component picture of And~II, the MDFs shown in the left and
right panels of Figure~11 are expected to predominantly sample the
spatially extended and exponential components, respectively. However,
neither will give a clean sample of that component.

To derive the {\it intrinsic} MDFs for the extended and exponential
components, we note that any MDF which samples some radial range of
And~II will be a weighted sum of the two intrinsic MDFs, $[MDF]_A$ and
$[MDF]_B$, such that

\begin{equation}
MDF = x [MDF]_A / (1 + x) + [MDF]_B / (1 + x)
\end{equation}

\noindent where $x$ is the ratio of the number of stars in component A
to component B, and is found by integrating the radial profiles of the
two components in the appropriate radial ranges. If we produce two
MDFs (using the same age assumption) for two different radial ranges in
And~II, then we can solve these equations to find $[MDF]_A$ and
$[MDF]_B$ (for that age assumption).

Figure~12 shows the {\it intrinsic} MDFs for the spatially extended
component (left panel) and the exponential component (right panel)
using this technique (under the age assumptions derived earlier). The
dashed histograms show the MDFs for the outer and inner regions
calculated previously. The MDF for the extended component is
unsurprisingly very similar to the MDF for the outer regions, given
the complete dominance of this component at large radius. On the other
hand, the MDF for the exponential component is significantly different
to the MDF for the inner regions given the non-negligible contribution
of the extended component in this radial range. The MDF for the
exponential component implies a mean metallicity of [Fe/H] $\simeq
-1$, in reasonable agreement with the estimate from the colour of the
red clump. Despite being significantly metal-rich, this component
nevertheless appears to possess a large tail to very low
metallicity. Interpretation of this feature is difficult, however,
since age-metallicity degeneracies will undoubtedly have an effect,
and the age of this component means that there could be contributions
from Asymptotic Giant Branch stars masquerading as bluer RGB stars.

We conclude that, in the two component model of And~II, the spatially
extended component is old ($\sim 13$\,Gyrs), has a mean metallicity of
[Fe/H] $\simeq -1.5$, and a dispersion of $\sigma_{{\rm [Fe/H]}} =
0.28$\,dex. The exponential component, on the other hand, is younger
($\sim 7 - 10$\,Gyrs) and more metal-rich ([Fe/H] $\simeq -1$) with
what appears to be a significant tail to low metallicities.

\section{Discussion}

\begin{table}
\begin{tabular}{rcc}
{}                                & Old               & Intermediate age \\
{}ANDROMEDA II                    & (extended)        & (concentrated)   \\
\hline\\
$n$                               & $0.3 \pm 0.1$     & $1.0 \pm 0.3$    \\
$r_o$ (arcmins)                   & $10.0 \pm 8.5$    & $2.2 \pm 0.8$    \\
$r_o$ (pc)                        & $1897 \pm 1612$   & $417 \pm 152$    \\
$\%$ light                        & $\sim 74$         & $\sim 26$   \\
$M_V$                             & $-12.3^{+0.3}_{-0.2}$ & $-11.2^{+0.3}_{-0.2}$\\
$L_V$ ($L_\odot \times 10^6$)     & $6.9^{+1.5}_{-1.2}$ & $2.5^{+0.6}_{-0.5}$\\
$\mu_{0,V}$ (mags\,arcsec$^{-2}$) & $\sim 26.1$       & $\sim 24.8$      \\
$\bar{t}$ (Gyrs)                  & $\sim 13$\,Gyrs   & $\sim 7 - 10$\,Gyrs   \\
${\rm[Fe/H]}$                     & $-1.5 \pm 0.1$    & $\sim 1.0$       \\
$\sigma_{\rm[Fe/H]}$              & $0.28$\,dex       & $0.40$\,dex      \\
\hline\\
\end{tabular}
\caption{Summary of properties for the two stellar components in Andromeda~II}
\end{table}

\subsection{Two components versus smoothly varying gradients}

And~II is a complex dwarf galaxy, both in terms of its structure and
its SFH. The inner region of the galaxy possesses
very different stellar populations to the outer region. Radial
gradients in the stellar populations of dwarf galaxies have been
observed many times before, most notably by \cite{harbeck2001}. Radial
gradients can be produced due to the presence of two or more
dynamically distinct stellar components with different stellar
populations. In this scenario, the gradient arises as a result of the
changing contributions of the two components as a function of
radius. Alternatively, gradients could arise within a single dynamical
component. For example, a centrally concentrated, chemically
homogeneous gas cloud will form stars everywhere; however, more stars
will be formed in the higher density central regions, which will
enrich the central parts of the cloud more than the outer
parts. Subsequent generations of stars at the center of the cloud will
therefore form out of more enriched gas than stars in the outer parts,
and so will evolve to occupy different loci in the colour-magnitude
diagram after a fixed time, thereby producing a gradient in the
stellar populations.

For the specific case of the population gradients in And~II, we
strongly favor the first scenario of multiple dynamical
components. Compelling evidence in favour of this interpretation comes
from the radial profiles shown in Figure~\ref{radial}. The radial
distributions of the red RGB and HB populations are dramatically
different, and effectively possess different functional forms; the
former is an exponential profile, whereas the latter is constant
density out to large radius. It is very unlikely that two populations
with such disparate spatial distributions could have formed out of the
same gas cloud and still possess the same kinematics. If both
populations are in equilibrium with the dSph then they cannot both
have the same kinematic properties while possessing such vastly
different density profiles and still satisfy Jeans equation. If
instead we argue that one or both of the populations is not in
equilibrium, then we have no reason to expect the two populations will
have the same kinematics.

In addition to these considerations, if we adopt the assumption that
And~II consists of two distinct components then the overall radial
profile of And~II must equal the sum of the radial profiles of its two
components. We find that a weighted sum of the HB and red RGB profiles
fits the overall radial profile of And~II very well. We also find that
the radius at which the central component begins to dominate is
coincident with the radius at which MI6 noticed their profile for
And~II deviated significantly from a single component fit. In the
Suprime-Cam data, the stellar populations inside this radius are very
different to the populations at larger radii, and have different
metallicities and mean ages. Given all of these factor, we conclude
that the change in the stellar populations of And~II as a function of
radius is due to the presence of two, dynamically distinct, stellar
components with distinct stellar populations.

Table~1 presents a summary of the main properties we derive for the
two components in And~II. \cite{dacosta2000} first showed that the
chemical abundance distribution of And~II could not be explained by a
single component simple chemical enrichment model and instead required
two contributing populations of different ages. The metal-poor
population had a mean metallicity of $\log\left(<Z>/Z_\odot\right) =
-1.6$ and the metal-rich population had $\log\left(<Z>/Z_\odot\right)
= -0.95$. The former outnumbered the latter by $\sim 2.3:1$. The
metallicity estimates we derive for the two components in And~II agree
well with the values implied by \cite{dacosta2000}; in addition, we
calculate the ratio of the metal-poor population to the metal-rich
population to be $\sim 2.8:1$ (assuming equal mass-to-light ratios),
in good agreement with their model. We also find that the younger,
more metal-rich population possesses a red clump; this is the first
time this feature has been observed in a M31 dSph companion. The M31
dSphs are often characterised as having no significant intermediate
age populations; we stress that these earlier results for And~II and
our own discovery of a red clump highlight that this characterisation
is probably inadequate.

\subsection{Andromeda II and the Local Group dwarf spheroidals}

\cite{tolstoy2004} were the first to show convincingly that the
Sculptor dSph consisted of two, spatially, chemically and dynamically
distinct populations. This is contrary to the prevailing treatment of
dSph galaxies as single stellar component systems supported by
velocity dispersion, embedded within a dark matter halo. Subsequently,
\cite{battaglia2006} have shown that the Fornax dSph consists of three
dynamically distinct populations of stars, and \cite{ibata2006} have
shown the presence of similarly distinct populations in Canes
Venatici. However, equivalent studies of Leo~I (\citealt{koch2006a}),
Leo~II (\citealt{koch2006b}) and Carina (\citealt{koch2006c}) have not
revealed any dynamically distinct stellar populations in these
galaxies. In all cases where dynamically distinct components and/or
radial gradients have been observed, the comparatively metal-rich
component is shown to be more centrally concentrated than the
metal-poor component.

And~II appears to bear some resemblance to Sculptor and Fornax. In all
these galaxies, the more concentrated populations are younger than the
more extended populations. \cite{kawata2006} examine the evolution of
the baryonic component of dwarf galaxies in a cosmological simulation
and show that they can reproduce this general feature as a result of
multiple episodes of star formation: later episodes of star formation
are spatially less extended than earlier episodes due to the
dissipative collapse of the gas component. In this model, the
populations form {\it in situ}, although this does not prevent the gas
which formed the younger populations from being recently accreted.

In contrast, \cite{battaglia2006} present tentative evidence for
non-equilibrium kinematics in the most extended (oldest) stellar
component in Fornax. This might suggest that its formation was driven
by mergers, perhaps by accreting an old dwarf galaxy or even a
globular cluster which was subsequently destroyed in the tidal field
of Fornax (see \citealt{coleman2004,coleman2005a} for tentative
evidence of past mergers in Fornax). \cite{ibata2006} also tentatively
propose that the two components in Canes Venatici are not in
equilibrium with the same dark matter halo. 

Other than an overall velocity dispersion for And~II based on 7 stars
(\citealt{cote1999a}), high quality global kinematics for And~II are
lacking.  It is tempting, however, to speculate on the origin of the
bizarre density profile of the extended component of And~II. This
component is effectively constant density out to large radius ($\sim
10$\,arcmins $\simeq 1.9$\,kpc) before truncating. This has not been
observed in any other dwarf galaxy to date and is in contrast to the
usual King, exponential, Plummer etc profiles of dwarf
galaxies. Interestingly, numerical simulations of the tidal disruption
of dwarf galaxies in MW-like potentials show some evidence of
producing debris which is distributed roughly evenly with radius
(eg. Figure~3 of \citealt{penarrubia2006}). Scaled down to dwarf
galaxy potentials, the same physics may be responsible here; that is,
the stars in the extended component of And~II may be the debris from
the accretion and eventual destruction of an old stellar system in
the potential of And~II. Thus perhaps this component is a relic of the
hierarchical formation of And~II. 

While intriguing, detailed simulations of the tidal destruction of
stellar systems in the potential of a dwarf galaxy do not yet exist to
study the above proposal in any detail. Indeed, a strong argument
against this scenario is that the extended component of And~II is the
oldest component and contributes the majority of the light. As such,
we would expect it to be the main body of And~II and not accreted
debris. The exponential component formed later, perhaps through a
mechanism similar to that proposed by \cite{kawata2006}. While this
does seem the most natural explanation, we are forced to address the
question of why the original body of And~II has such an unusual and
unique density profile. This is a fascinating question, and one which
we are currently unable to answer.

\section{Summary}

We have used Subaru Suprime-Cam to obtain multi-colour imaging of the
M31 dSph And~II to a depth equivalent to earlier HST-WFPC2 studies but
over an area 100\,times larger. We identify various stellar
populations, including a red clump. This is the first time this
feature has been detected in a M31 dSph, which are often characterized
as having no significant intermediate age populations. In the case of
And~II, this characterization is incorrect.

We construct radial profiles for the various stellar populations and
show that the HB has a nearly constant density spatial distribution
out to large radius, whereas the reddest RGB stars are centrally
concentrated in an exponential profile. We argue that these
populations trace two distinct structural components in And~II, and
show that this assumption provides a good match to the overall radial
profile of this galaxy. We go on to demonstrate that the two
components have very different stellar populations; the exponential
component has an average age of $\sim 9$\,Gyrs old, is relatively
metal-rich ([Fe/H] $\sim -1$) with a significant tail to low
metallicity, and possesses a red clump. The extended component, on the
other hand, is ancient, metal-poor ([Fe/H] $\sim -1.5$) with a
narrower dispersion $\sigma_{\rm [Fe/H]} \simeq 0.28$, and has a well
developed blue HB. 

The extended component contains approximately three-quarters of the
light of And~II. Its unusual surface brightness profile is unique in
Local Group dwarf galaxies, and implies that its formation and/or
evolution may have been quite different to most other dSphs. The two
component structure, however, is also observed in Fornax and Sculptor,
and it may be that the centrally concentrated exponential component
formed in the way envisioned by \cite{kawata2006} through the
dissipative collapse of gas after the main component had formed. The
chemo-dynamical structure of And~II is clearly very complex and
warrants spectroscopic studies of its metallicity and kinematic
properties. It lends yet further support to the accumulating body of
evidence which suggests that the evolutionary histories of faint dSph
galaxies can be as every bit as complicated as their brighter and more
massive counterparts.

\section*{Acknowledgements}
We thank Gary Da Costa for sharing his HST-WFPC2 data with us and his
assistance in cross correlating it with these Subaru data, and the
anonymous referee for many suggestions which lead to a significant
improvement in this paper. AWM thanks Jorge Pe{\~n}arrubia, Julio
Navarro, Kim Venn and Don VandenBerg for useful and enjoyable
conversations during the preparation of this paper, and S. Ellison and
J. Navarro for financial assistance. AWM is supported by a Research
Fellowship from the Royal Commission for the Exhibition of 1851.  This
work is partly supported by a Grant-in-Aid for Science Research
(No.16540223) by the Japanese Ministry of Education, Culture, Sports,
Science and Technology.

\bibliographystyle{apj}
\bibliography{/Users/Alan/Papers/references}

\end{document}